\documentstyle[preprint,aps,epsf]{revtex}

\tighten
\draft

\begin{document}
\title{{\bf Isospin dependence of liquid-gas phase transition in hot asymmetric
nuclear matter}}
\author{Wei Liang Qian$^1$  Ru-Keng Su$^{2.1}$  Ping Wang$^{1.3}$}

\address{$^{1}$Department of Physics, Fudan University, Shanghai 200433, P.R.China}
\address{$^{2}$China Center of Advanced Science and Technology (World Laboratory),\\
P. O. Box 8730, Beijing, P.R.China}
\address{$^{3}$Institute of High Energy Physics, P.O.Box 918, Beijing 100039,\\
P.R.China}
\maketitle

\begin{abstract}
By using the Furnstahl, Serot and Tang's model, the effect of density
dependence of the effective nucleon-nucleon-$\rho $-meson (NN$\rho $)
coupling on the liquid-gas phase transition in hot asymmetric nuclear matter
is investigated. A limit pressure p$_{\lim }$ has been found. We found that the
liquid-gas phase transition cannot take place if $p>p_{\lim }$. The binodal
surface for density dependent NN$\rho $ coupling situation is addressed.
\end{abstract}


It is generally recognized that the liquid-gas (LG) phase transition of one
component system is of first order. The chemical potential continues at the
phase transition point but its first order derivatives, namely, entropy and
volume, are discontinuous. But for a multi-components or multi-conserved
charges system, as was pointed out by M\H{u}ller and Serot \cite{1}, because
of the greater dimensionality of the binodal surface, the LG phase
transition can be of second order, i.e., the entropy continues but the
second order derivatives of chemical potential (for example, capacity) are
discontinuous. An asymmetric nuclear matter has two components of proton and
neutron, and two conserved charges of baryon number and the third component
of isospin, will undergo a continuous second order phase transition.

Obviously, because of charge independence, the basic difference between
proton and neutron be isospin. The isospin dependent interactions of
nucleon-nucleon-isovector mesons play the key role to address the LG phase
transition. As was pointed out by our previous papers \cite{2,3,4}, if one
employed the isospin independent model, for example, Welacka model \cite{5}
or Zimanyi-Moszkowski model \cite{6}, to investigate asymmetric nuclear
matter, a lot of difficulties, e.g. Coulomb instability and negative
asymmetric parameter in the vapor phase, will emerge. To overcome these
difficulties, the isospin vector $\rho $-meson must be introduced. It can be
shown that the chemical potentials of proton and neutron may depend on the
third component of isospin when NN$\rho $ interaction exists. A model
without isospin vector $\rho $-meson, or even if it has $\rho $-meson, but
the chemical potentials of the proton or neutron are still independent of
the NN$\rho $ interaction because the third component I$_3$ of isospin be
zero such as in a symmetric nuclear matter, the LG phase transition is still
of first order.

In fact, the chemical potentials of proton and neutron not only depend on I$
_3$ but also on the effective NN$\rho $ coupling g$_\rho $. Then the
effective coupling g$_\rho $ is also essential for studying the LG phase
transition because the chemical potentials determine the binodal surface
directly. In our previous papers \cite{7,8,9}, we have shown that the effective
couplings depend on the density and temperature in a hot and dense nuclear
matter. By using the Thermo Field Dynamics \cite{5} to calculate the
three-lines vertices Feynman diagrams of NN$\pi $, NN$\sigma $ , NN$\omega $
and NN$\rho $ interactions, we have found that the effective couplings of g$
_\pi $, g$_s$, g$_v$ and g$_\rho $ all decrease as the nucleon density
increases. In an asymmetric nuclear matter, one can easily prove that the
chemical potentials of nucleons have a term which is proportional to g$_\rho
^2$ and I$_3$. This term has opposite signs for proton and neutron due to
their different third component of isospin. Obviously, if g$_\rho $ depends
on density, this term will change and then the chemical potentials of proton
and neutron, as well as the binodal surface of LG phase transition will also
be changed. The objective of the present paper is to investigate the effect
of the density dependence of g$_\rho $ on LG phase transition. We will prove
that if g$_\rho $ is a decreasing function of density, a limit pressure $p
_{\lim }$ will occur, when $p>p_{\lim }$, the coexistenced equations have
no solution and the LG phase transition cannot be existed. The chemical
potential of neutron will become a monotoneous functioin of asymmetry $
\alpha $ in this case.

To illustrate our result, we employ a model suggested by Furnstahl, Serot
and Tang (FST) \cite{11,12,13,14} recently. This model is an extension of
quantum hadrodynamics and has been proven to be successful to explain many
experimental properties of both nuclear matter and the finite nuclei in mean
field approximation. The Lagrangian density of FST model under mean field
approximation is 
\begin{eqnarray}
L_{MFT} &=&\overline{\Psi }\left[ i\gamma ^\mu \partial _\mu -\left(
M-g_s\phi _0\right) -g_v\gamma ^0V_0-\frac 12g_\rho \tau _3\gamma
^0b_0\right] \Psi  \label{e1} \\
&&+\frac 12m_v^2V_0^2\left( 1+\eta \frac{\phi _0}{S_0}\right) +\frac 1{4!}
\zeta \left( g_vV_0\right) ^4+\frac 12m_\rho ^2b_0^2  \nonumber \\
&&-H_q\left( 1-\frac{\phi _0}{S_0}\right) ^{4/d}\left[ \frac 1dln\left(
1-\frac{\phi _0}{S_0}\right) -\frac 14\right]  \nonumber
\end{eqnarray}
where g$_s$, g$_v$ g$_\rho $ are, respectively, the couplings of light
scalar meson $\sigma $, vector meson $\omega $ and isovector meson $\rho $
fields to the nucleon, $\phi _0$, V$_0$, b$_0$ are the expectation values $
\phi _0\equiv <\phi >$, $<V_\mu >\equiv \delta _{\mu 0}V_0$, $<b_{\mu
3}>\equiv \delta _{\mu 0}b_0$. The scalar fluctuation field $\phi $ is
related to $S$ by $S\left( x\right) =S_0-\phi \left( x\right) $ and $H_q$ is
given by $m_s^2=4H_q/\left( d^2S_0^2\right) $, d the scalar dimension. By
using the standard technique of statistical mechanics, we get the
thermodynamic potential $\Omega $ as \cite{15} 
\begin{eqnarray}
\Omega &=&V\left\{ H_q\left[ \left( 1-\frac{\phi _0}{S_0}\right)
^{4/d}\left( \frac 1dln\left( 1-\frac{\phi _0}{S_0}\right) -\frac 14\right) +
\frac 14\right] \right.  \label{e2} \\
&&\left. -\frac 12m_\rho ^2b_0^2-\frac 12\left( 1+\eta \frac{\phi _0}{S_0}
\right) m_v^2V_0^2-\frac 1{4!}\zeta \left( g_vV_0\right) ^4\right\} 
\nonumber \\
&&-2k_BT\left[ \sum_{k,\tau }ln\left( 1+e^{-\beta \left( E^{*}\left(
k\right) -\nu _\tau \right) }\right) +\sum_{k,\tau }ln\left( 1+e^{-\beta
\left( E^{*}\left( k\right) +\nu _\tau \right) }\right) \right]  \nonumber
\end{eqnarray}
where $\beta =1/k_BT$ and the quantity $\nu _i$ $\left( i=n,p\right) $ is
related to the usual chemical potential $\mu _i$ by the equations 
\begin{equation}
\nu _n=\mu _n-g_vV_0+\frac{g_\rho ^2\rho _3}{4m_\rho ^2}  \label{e3}
\end{equation}
\begin{equation}
\nu _p=\mu _p-g_vV_0-\frac{g_\rho ^2\rho _3}{4m_\rho ^2}  \label{e4}
\end{equation}
where $\rho _3=\rho _p-\rho _n$ and the third component of isospin I$
_3=\left( N_p-N_n\right) /2=V\rho _3/2$. The third term of the right hand
side of Eq(\ref{e3}) and Eq(\ref{e4}) depends on $\rho _3$ and g$_\rho ^2$.
They have opposite signs and play the essential role to determine the LG
phase transition.

Having obtained the thermodynamic potential, all other thermodynamic
quantities, for example, pressure p=$-\Omega /V$, can be calculated. The
two-phase coexistence equations are 
\begin{equation}
\mu _i^L\left( T,\rho _i^L\right) =\mu _i^V\left( T,\rho _i^V\right)
\label{e5}
\end{equation}
\begin{equation}
p^L\left( T,\rho _i^L\right) =p^V\left( T,\rho _i^V\right)  \label{e6}
\end{equation}
where subscripts of one phase L and V stand for liquid and vapor,
respectively. The stability conditions are given by \cite{1} 
\begin{equation}
\rho \left( \frac{\partial p}{\partial \rho }\right) _{T,\alpha }=\rho
^2\left( \frac{\partial ^2{\cal F}}{\partial \rho ^2}\right) _{T,\alpha }>0
\label{e7}
\end{equation}
\begin{equation}
\left( \frac{\partial \mu _p}{\partial \alpha }\right) _{T,p}<0\text{ or }
\left( \frac{\partial \mu _n}{\partial \alpha }\right) _{T,p}>0  \label{e8}
\end{equation}
where ${\cal F}$ is the density of free energy, $\alpha =\left( \rho _n-\rho
_p\right) /\rho $ the asymmetric parameter, and $\rho =\rho _n+\rho _p$.

The numerical calculatioins have been done by adopting the parameters set T
1 of FST model \cite{11,12,13}\.{ }The parameters of set T1 are 
\begin{eqnarray}
g_s^2 &=&99.3,\hspace{5mm}g_v^2=154.5,\hspace{5mm}g_\rho ^2=70.2  \label{e9}
\\
m_s &=&509MeV{\bf ,}\hspace{5mm}S_0=90.6MeV  \nonumber \\
\zeta  &=&0.0402,\hspace{5mm}\eta =-0.496,\hspace{5mm}d=2.70  \nonumber
\end{eqnarray}
Our results for g$_\rho =\left( 70.2\right) ^{1/2}$=constant are shown in
Fig.1 and Fig.2 by solid curves. The Gibbs conditions (\ref{e5}) and (\ref
{e6}) for phase equilibrium demand equal pressures and chemical potentials
for two phase with different concentrations. The collection of all such
pairs $\alpha _1\left( T,p\right) $ and $\alpha _2\left( T,p\right) $ form
the binodal suface. In Fig.1, the chemical isobar vs. $\alpha $ curves at
fixed temperature T=10MeV{\bf \ }and p=0.100MeV$\left( fm\right) ^{-3}$ are
labeled by A and A' for neutron and proton respectively. The two desired
solutions form the edges of a rectangle and can be found by means of the
geometrical construction shown in Fig.1 \cite{1}. The critical curves with
T=10MeV{\bf \ }and p$_{crit}=0.165$MeV$\left( fm\right) ^{-3}$ are shown in
Fig.1 by B and B' where the chemical potential curve arrive at a inflection
point and the rectangle is degenerate to a line vertical to the $\alpha $
axis. The behaviour of the nuclear matter under isothemal compression, or in
other words, the section of binodal surface at finite temperature T=10MeV
{\bf \ }are shown in Fig.2. The physical behaviour of this processes has
been discussed by ref.\cite{1}. Assume that the system is initially prepared
with $\alpha =0.6$ (gas), during the compression, the two-phase region is
encountered at point A, and the liquid phase emerges at point B. The gas
phase evolves from A to D, while the liquid phase evolves from B to C. The
system leaves the region of instability at point C, while the original gas
phase is about to disappear. The critical point (CP), the point of equal
concentration (EC) and the maximal asymmetry (MA) are indicated in Fig.2.

Now we are in a position to extend our discussion to the case of g$_\rho $
density dependence. In fact, the effective masses of nucleons, effective
masses or screening masses of mesons, and the effective coupings of
NN-mesons are all dependent on density and temperature. We can sum the
tadpole diagrams and the exchange diagrams for nucleon, the vacumm
polarization diagrams for mesons and the three-lines vertex diagrams for
effective couplings to get their density and temperature dependence \cite
{4,7,8,9,15,16,17}. But in order to illustrate our result more
transparently, instead of the exact calculation of three-lines vertex, we
introduce an {\it ansatz} 
\begin{equation}
g_\rho ^{\prime }=g_\rho \left[ 1-A\rho +B\rho ^2\right]  \label{e10}
\end{equation}
where A, B are two adjust parameters. The reason for our choice are: at
first, the three lines vertex calculations are model dependent, but we hope
that our investigation can be more general; secondly adjust the values of A,
B can made $g_\rho ^{\prime }$ be decreased or increased with density, then
we can study the LG phase transition for two different cases. We can imagine
that the Eq(\ref{e10}) is an density expansion of effective coupling $g_\rho
^{\prime }$ at low density regions.

The results for density dependent $g_\rho ^{\prime }$ ansatz Eq(\ref{e10})
are shown in Fig.3, Fig.4 and Fig.5. We see from Fig.3 and Fig.4 that the
chemical potential of neutron $\mu _n$ increases rapidly with density. It
passes through an inflection point and becomes monotoneous when pressure
increases. But the shape of $\mu _p$ vs. $\alpha $ curves change slowly. Then
we will get a limit pressure p$_{\lim }$, when $p>
p_{\lim }$, the rectangle cannot be found and the coexistenced equations
have no solution. The last rectangle in the chemical isobar vs. $\alpha $
curves for A=1, B=0, T=10MeV, and p$_{\lim }$=0.130MeV$\left( fm\right)
^{-3} $ is shown in Fig.3 by dashed lines, where $\alpha _1$=0.62 and $
\alpha _3$=0.75 correpond to the maximum and the minimum of $\mu _n$
respectively. The pair $\alpha _1$=0.62 and $\alpha _2$=0.84 form the end of
the binodal surface, as shown in Fig.5. The curve for A=1, B=0, T=10{\bf \ }
MeV{\bf ,} but p=0.145MeV$\left( fm\right) ^{-3}$ ($p>
p_{\lim }$) is shown in Fig.4. We see that $\mu _n$ becomes
monotoneous at this pressure, and no rectangle can be found. The relation
between limit pressure and parameters A and B for a fixed temperature T=10
MeV{\bf \ }is shown in Table 1. We find from Table 1 when A and B increase,
the effective coupling $g_\rho ^{\prime }$ decreases and the limit pressure
decreases. If A changes its sign to become negative, $g_\rho ^{\prime }$
will increase with density, and in this case, instead of $\mu _n$ , $\mu _p$
becomes monotoneous. The limit pressure is still existed but decrease when A
and B decrease.

The section of binodal surface for A=1, B=0, T=10MeV{\bf \ }is shown in
Fig.5. We see from Fig.5 that the curve will cut off at limit temperature p$
_{\lim }$ clearly. The total asymmetric parameter $\alpha $ is divided into
four regions, namely, $\left[ 0,\alpha _1\right] $, $\left[ \alpha _1,\alpha
_3\right] $, $\left[ \alpha _3,\alpha _2\right] $ and $\left[ \alpha
_2,\alpha _{\max }\right] $. The physical behaviour of isothermal
compression in different regions are different. We find:

(1) If the system is initially prepared with $0<\alpha  <\alpha _1$, the
precess of isothermal compression is similar to that of the case with
constant $g_\rho $. It begin at gas phase, suffers a second order LG phase
transition and ends at liquid phase.

(2) If the initial $\alpha $ is located at the region $\alpha _2 <\alpha <
\alpha_{max }$, the system enters and leaves the two-phase region on the
same branch, so the system remain in the same gase phase. As was pointed out
by M\H{u}ller and Serot \cite{1}, this retrograde condensation is unique to
the binary system and dose not occur in one-component systems.

(3) Suppose that the initial $\alpha $ is located at the region $\alpha _1
<\alpha <\alpha _3$. Since $\alpha _1$ and $\alpha _3$ correspond to the
maximum and minimum of $\mu _n$, we find $\left( \partial \mu _n/\partial
\alpha \right) _{T,p}<0$ in this region and the stability condition Eq.(\ref
{e8}) will be destroied. The system begins at gas phase, enters a two-phase
region and becomes instable at the limit pressure.

(4) If the initial $\alpha $ is located at the region $\alpha _3<\alpha <
\alpha _2$, the behaviou of the system is similar to that of the case (3),
except it will be ended to a stable phase at the limit pressure because the
stability condition $\left( \partial \mu _n/\partial \alpha \right) _{T,p}>0$
is satisfied.

In summary, we have shown that the density dependence of effective NN$\rho $
coupling is important to the LG phase transition. A limit pressure p$_{\lim
} $ has been found for a fixed temperature and the LG phase transition
cannot take place in asymmetric nuclear matter provided $p>p_{\lim }$ . Of course,
for a fixed pressure, we can also get a limit temperature. This conclusion
is similar to that of the Coulomb instability \cite{2,3,4} of nuclei. The
basic difference is that instead of finite nuclei, our conclusion has be
found to be suitable for asymmetric nuclear matter. Finally, we would like
to emphasize that the isospin is very important for the LG phase transition
of asymmetric nuclear matter.

The work was support in part by the National Natural Science Foundation of 
China under contract no. 19975010,
and the Foundation of Education Department of China.

\begin{table}[h]
\caption{relation between the limit pressure and adjust parameter A, B}
\begin{tabular}{llllllll}
$A$ & 1 & 2 & 3 & 5 & -2 & -2 & -5 \\ 
\hline
$B$ & 0 & 1 & 1 & 2 & 1 & -1 & -2 \\ 
\hline
$p_{\lim }{(MeV\left( fm\right) ^{-3})}$ & 0.13 & 0.115 & 0.105 & 0.90 & 0.125 & 0.125 & 0.115 \\ 
\end{tabular}
\end{table}

\begin{figure}[h]
\caption{ The chemical isobar as a function of $\alpha $ at fixed
temperature T=10MeV and the geometrical construction used to obtain the
asymmetry parameters in the two coexisting phases. B, B' correspond to the
critical curves. }
\label{Fig1}
\end{figure}

\begin{figure}[h]
\caption{ The section of binodal surface at T=10MeV, where CP, EC, and MA
stand for critical point, equal concentratioin and maximal asymmetry
respectively. }
\label{Fig2}
\end{figure}

\begin{figure}[h]
\caption{ The chemical isobar as a function of $\alpha $ for A=1, B=0, T=10
MeV and p=0.145 MeV$\left( fm\right) ^{-3}$. }
\label{Fig3}
\end{figure}

\begin{figure}[h]
\caption{ The chemical isobar as a function of $\alpha $ for A=1, B=0, T=10
MeV and p=0.13 MeV$\left( fm\right) ^{-3}$. }
\label{Fig4}
\end{figure}

\begin{figure}[h]
\caption{ The section of binodal surface for A=1, B=0, T=10MeV. $\alpha _1$, 
$\alpha _2$ correspond to the maximum and minimum of $\mu _n$ respectively. }
\label{Fig5}
\end{figure}

\begin{figure}[h]
\epsfbox{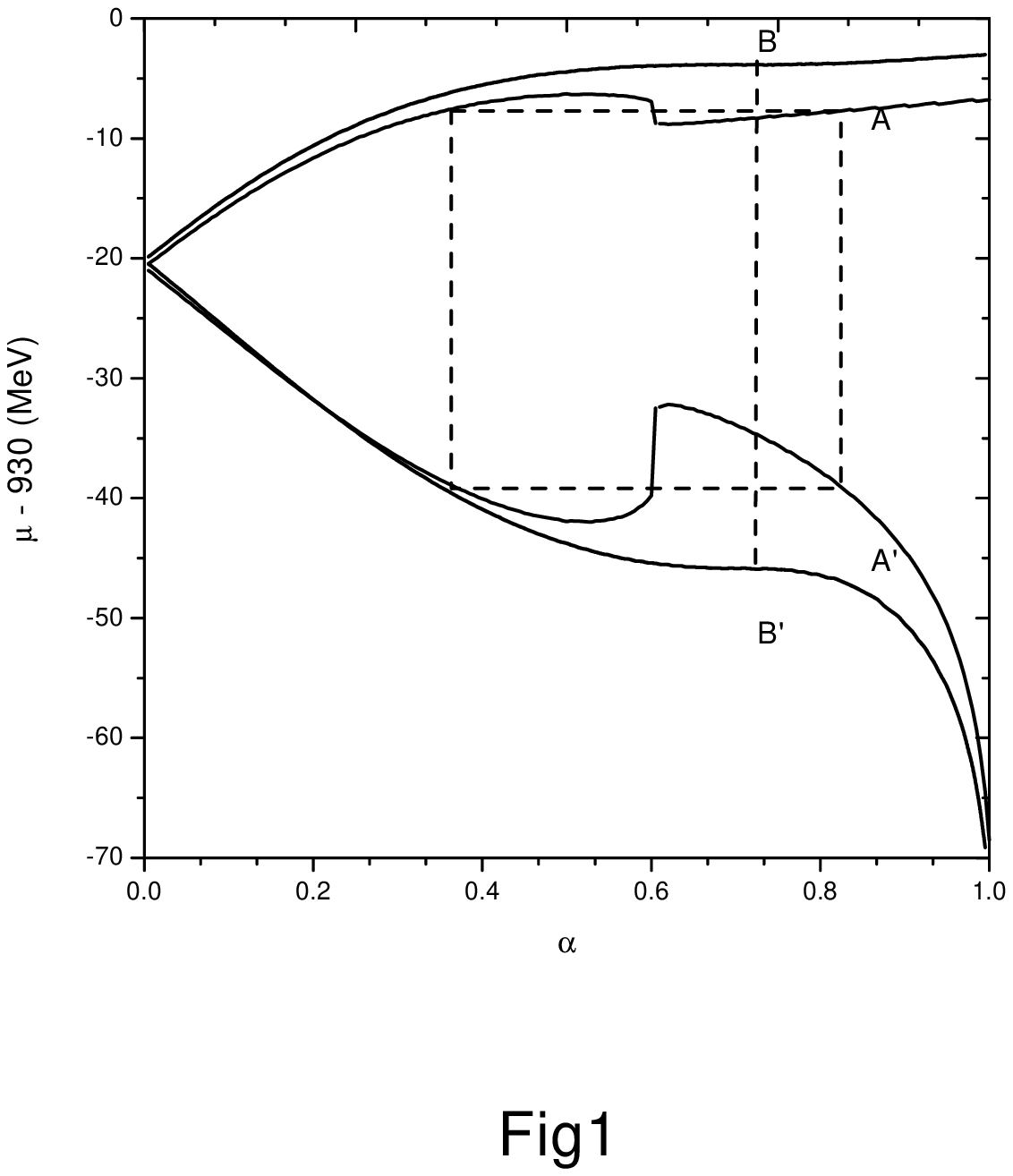}
\end{figure}

\begin{figure}[h]
\epsfbox{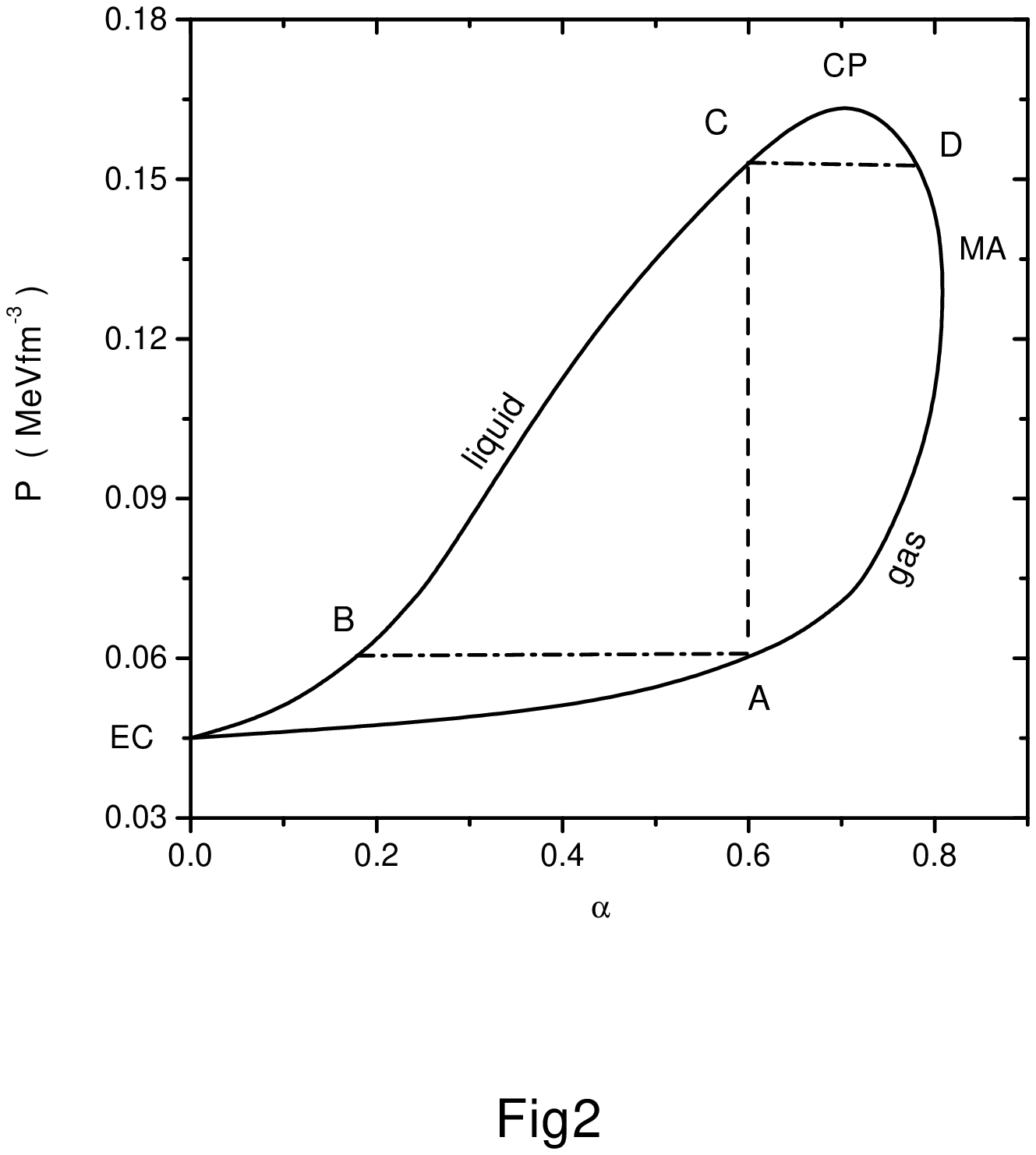}
\end{figure}

\begin{figure}[h]
\epsfbox{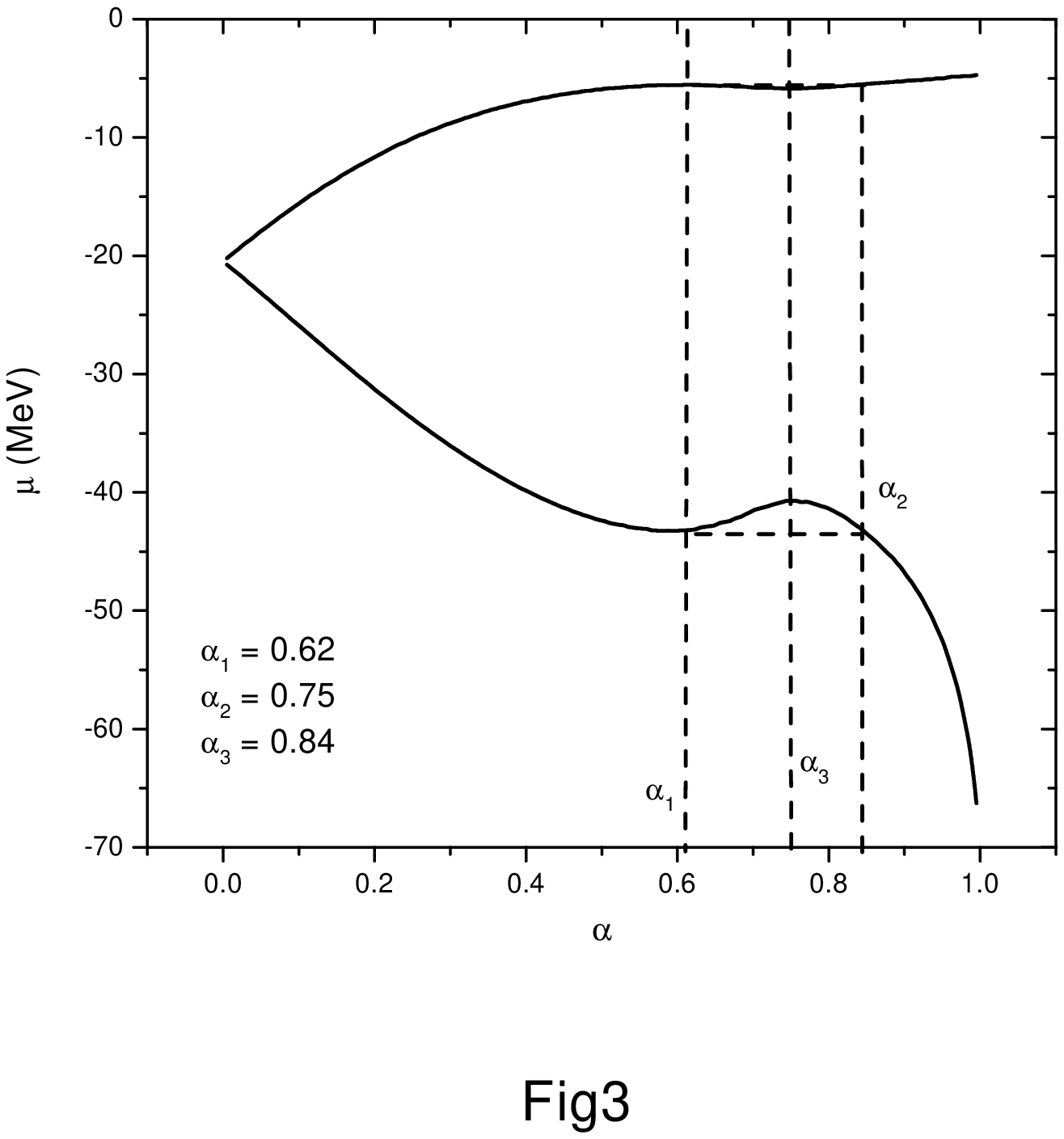}
\end{figure}

\begin{figure}[h]
\epsfbox{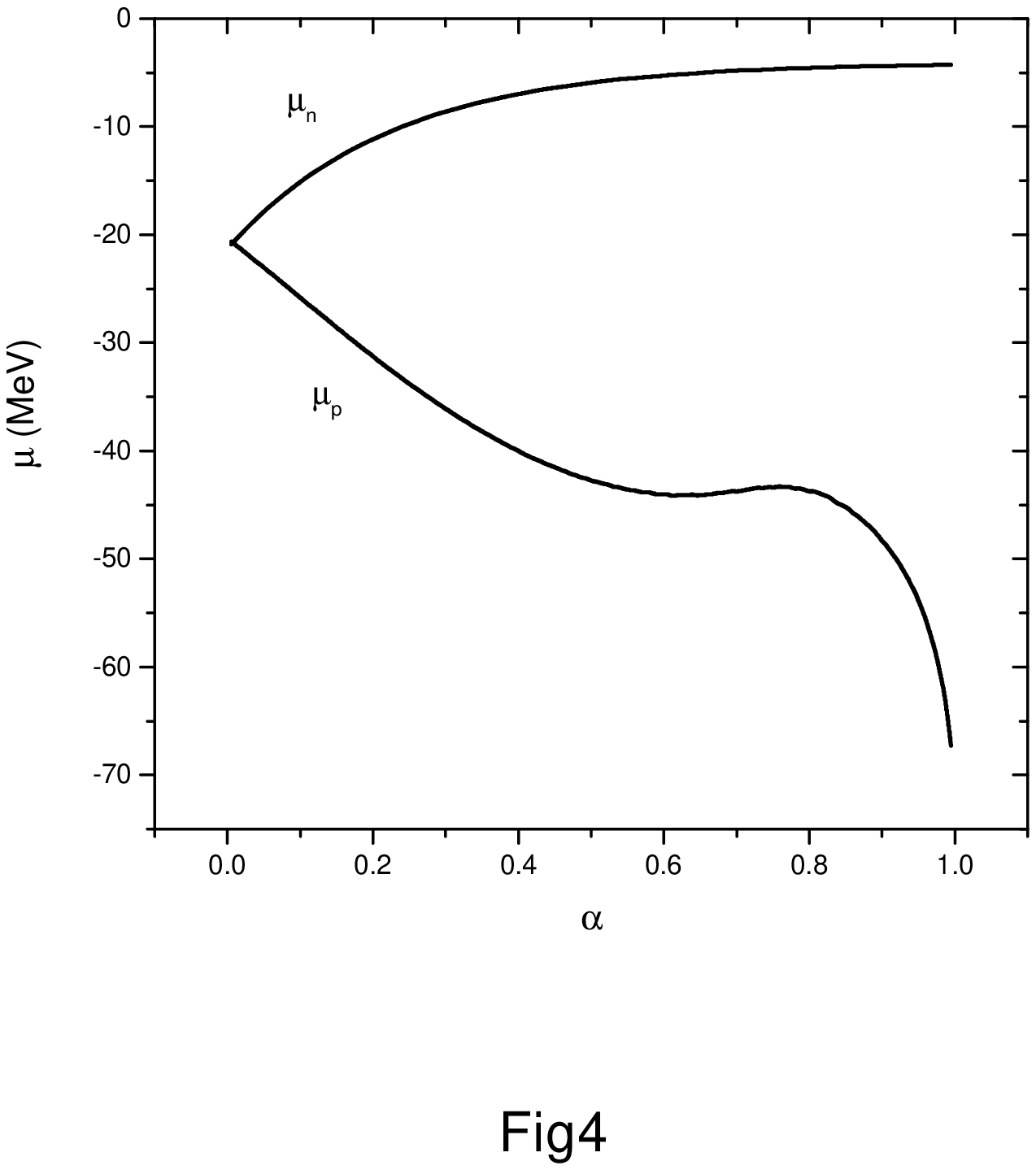}
\end{figure}

\begin{figure}[h]
\epsfbox{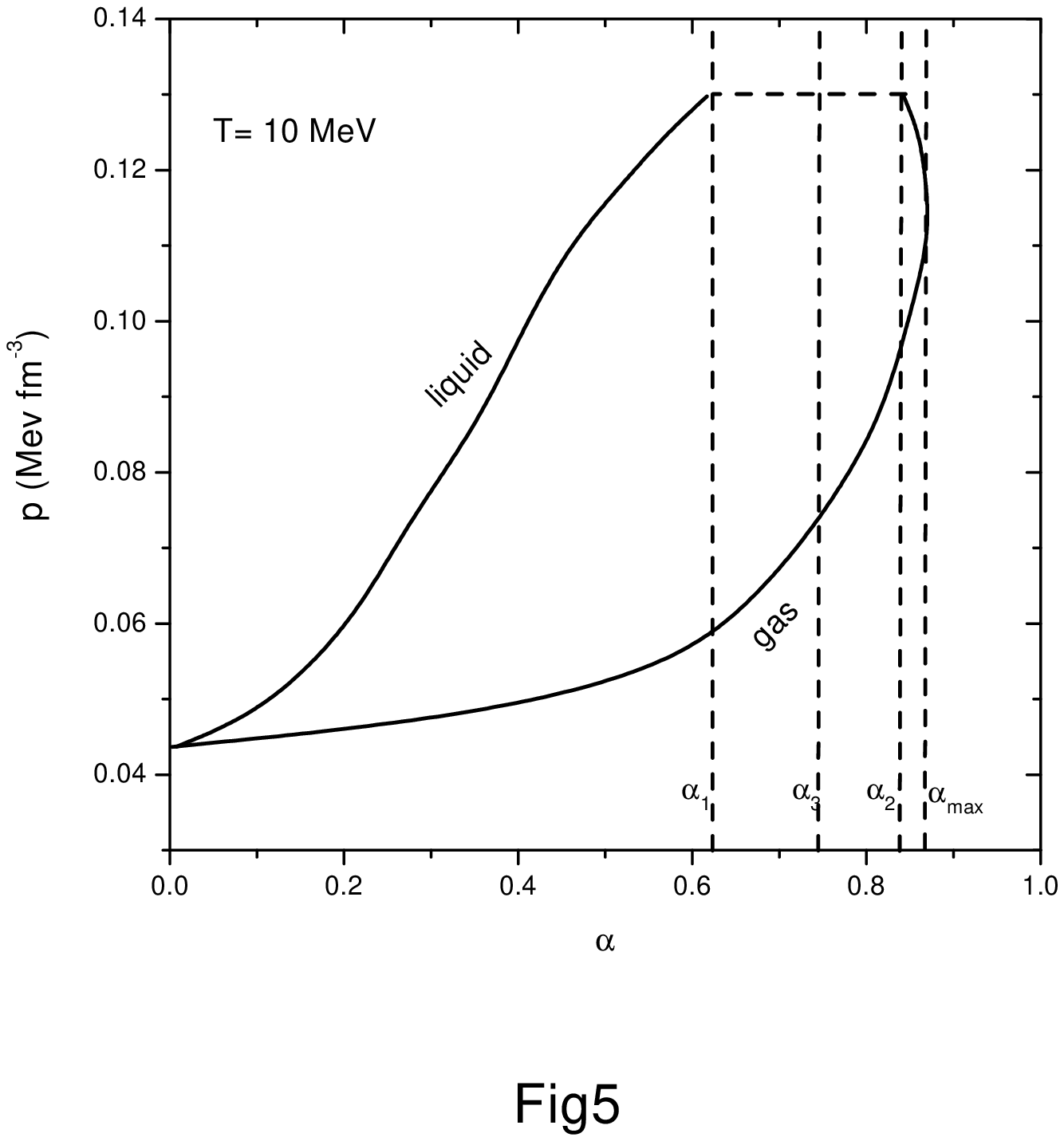}
\end{figure}


\begin{references}
\bibitem{1}  H.M\H{u}ller and B.D.Serot, Phys. Rev. C52 2072 (1995)

\bibitem{2}  H.Q.Song, Z.X.Qian and R.K.Su, Phys. Rev. C47 2001(1993)

\bibitem{3}  H.Q.Song, Z.X.Qian and R.K.Su, Phys. Rev. C49 2924(1994)

\bibitem{4}  P.Wang, Z.W.Chong, R.K.Su and P.K.N.Yu Phys. Rev. C59 928(1999)

\bibitem{5}  B.D.Serot and J.D.Walecka, Advances in Nuclear Physics, edited
by J.W.Negels and E.Vogt Plenam, New York, 1986), Vol. 16. P.1

\bibitem{6}  J.Zimanyi and J.A.Moszkowski, Phys. Rev. C42 1416 (1990)

\bibitem{7}  S.Gao, Y.J.Zhang and R.K.Su, Nucl. Phys. A593 362 (1995)

\bibitem{8}  Z.X.Qian, C.G.Su and R.K.Su, Phys. Rev. C47 877 (1993)

\bibitem{9}  R.K.Su, G.T.Zheng and G.G.Siu, J.Phys. G19 79 (1993)

\bibitem{10}  H.Umezawa, H.Matsumoto and M.Tachiki, Thermo Field Dynamics
and Condensed States (North-Holland, Amsterdam, 1982)

\bibitem{11}  R.J.Furnstahl, H.B.Tang and B.D.Serot, Phys. Rev. C52 1368
(1995)

\bibitem{12}  R.J.Furnstahl, B.D.Serot and H.B.Tang, Nucl. Phys. A598 539
(1996)

\bibitem{13}  R.J.Furnstahl, B.D.Serot and H.B.Tang, Nucl. Phys. A615 441
(1997)

\bibitem{14}  B.D.Serot and J.D.Walecka, Int. Jour. of Mod. Phys. E6 515
(1997)

\bibitem{15}  L.L.Zhang, H.Q.Song, P.Wang and R.K.Su, Phys. Rev. C59 3292
(1999)

\bibitem{16}  Y.J.Zhang, S.Gao and R.K.Su, Phys. Rev. C56 3336 (1997)

\bibitem{17}  S.Gao, Y.J.Zhang and R.K.Su Phys. Rev. C53 1098 (1996)
\end{references}
\end{document}